\documentclass[]{aa}  

\usepackage{tikz}
\usetikzlibrary{arrows.meta,positioning,decorations.markings}
\usetikzlibrary{arrows.meta,positioning}
\usetikzlibrary{arrows.meta,decorations.pathmorphing,positioning}
\usetikzlibrary{arrows.meta,positioning,decorations.pathreplacing}

\usepackage{float}

\usepackage{graphicx}
\usepackage{subfigure}
\usepackage{adjustbox}
\usepackage{txfonts}
\usepackage{natbib}
\usepackage{hyperref}

\begin{document}

   \title{Tidal Tails of Galactic Open Clusters: A Review of Methods,
Evidence, and Sources of Disagreement}

  \author{T.~Ramezani\inst{1}
  \and J.~Balčir\'akov\'a\inst{1,2}
	 \and P.~Mondal\inst{1}
	 \and K.~Neumannov{\'a}\inst{1}
  \and E.~Paunzen\inst{1}
	 \and G.~Sz{\'a}sz\inst{1}}

   \institute{Department of Theoretical Physics and Astrophysics, Faculty of Science, Masaryk University, Kotl\'{a}\v{r}sk\'{a} 2, 611 37 Brno, Czechia
   \email{Taherehramezani7@gmail.com}
    \and CESNET, Gener\'{a}la Píky 430/26 160 00 Prague 6, Czechia}
   \date{}

\date{}
 
  \abstract
{Tidal tails are a natural consequence of the dynamical evolution of open clusters in the Galactic tidal field, but their observational identification is intrinsically difficult because their surface density is low and the Galactic disc contains a dense and kinematically structured field population.}
{This review paper provides a critical synthesis of the observational and theoretical status of tidal tails of Galactic open clusters and, in particular, assesses how membership definitions, data dimensionality, clustering algorithms, and dynamical assumptions affect reported tail detections.}
{We compare the principal approaches used with Gaia data, including convergent-point and co-moving selection, density-based clustering, probabilistic membership models, machine-learning methods, and orbit/N-body-based filtering. We compile published measurements for a representative sample of nearby clusters and explicitly distinguish robust detections from candidate or method-dependent extended structures.}
{The literature has progressed from detailed studies of a few nearby clusters to systematic searches involving tens to hundreds of systems. The Hyades remains a benchmark because independent analyses recover coherent leading and trailing tails, but the extent and membership of the tails depend on the adopted selection. Large Gaia DR3 studies demonstrate that extended populations are common among nearby and intermediate-age clusters, while also showing that some structures can be produced or exaggerated by projection, field contamination, or the assumptions built into the membership model. Recent homogeneous analyses report tails in most of the nearby clusters in selected samples, but the corresponding catalogues should not be interpreted as equivalent because their completeness, contamination, and physical definitions of a ``tail member'' differ.}
{A useful synthesis is therefore not a binary catalogue of clusters with or without tails. Instead, tidal structures should be described using a hierarchy of evidence: astrometric coherence, photometric consistency, chemical or radial-velocity confirmation, orbital consistency, and agreement with dynamical models. Future progress will require common validation criteria and joint Gaia, spectroscopic, deep-photometric, and N-body analyses.}


   \maketitle
%

\section{Introduction} \label{introduction} 

Open clusters are gravitationally bound stellar systems embedded in the Galactic disc. Their subsequent evolution is governed by internal two-body relaxation, stellar evolution, binary interactions, and the external Galactic tidal field. As stars diffuse through phase space, a fraction of them reach escape conditions and leave the cluster. The escaped stars do not immediately become indistinguishable from the field: for a finite time, they retain correlated positions and velocities and can form elongated leading and trailing structures around the cluster orbit. These structures are commonly referred to as tidal tails, although the observational literature also uses terms such as corona, halo, extended structure, stream, or extra-tidal population for related but not necessarily identical phenomena \citep{Kupper2010}.
\\
The physical picture is now well established. In a rotating frame centred on a cluster, the Galactic tidal field produces two preferred escape channels near the inner and outer Lagrange points. Escapers acquire small offsets in orbital energy and angular momentum and subsequently execute epicyclic motion around the cluster orbit. Even in a smooth and time-independent Galactic potential, this process can generate density variations along the tails \citep{Just2009}. Time-dependent perturbations, including passages through the Galactic disc or spiral structure and encounters with giant molecular clouds, can further modify the escape rate and produce asymmetries or clumps \citep{Ernst2011}.
\\
Tidal tails are therefore useful in two complementary senses. First, they trace the dynamical dissolution of the cluster and provide information on mass loss, mass segregation, and the transition from a bound cluster to the field population. Second, their orientation and phase-space structure contain information about the cluster orbit and the Galactic potential \citep{Kupper2010}. The interpretation is nevertheless non-trivial because a tidal tail is not defined only by a projected overdensity. A convincing tail member should ideally be spatially extended from the core, kinematically compatible with the cluster, photometrically consistent with its stellar population, and dynamically plausible as an escaper.
\\
Before Gaia, this combination of requirements was rarely achievable for open clusters. The low surface density of an open-cluster tail is particularly problematic in the Galactic disc, where unrelated field stars are numerous, and the background is structured in both position and velocity. Gaia changed the situation by providing homogeneous all-sky astrometry for a very large stellar sample. Gaia DR2, EDR3, and DR3 successively improved the precision and completeness of parallaxes and proper motions and made it possible to search for extended cluster populations in five- and, for a subset of stars, six-dimensional phase space \citep{GaiaCollaboration2018,GaiaCollaboration2021,GaiaCollaboration2023}.
\\
The first Gaia-era studies demonstrated the potential of this approach. The Hyades tails were detected independently using a modified convergent-point analysis and a six-dimensional phase-space search \citep{Roser2019}. Subsequent studies extended the investigation to Praesepe, Coma Berenices, Blanco 1, NGC 2516, NGC 752, IC 4756, M\,67, NGC 2506, and other systems \citep{Ye2021}. Large catalogues subsequently demonstrated that extended stellar populations are common among nearby open clusters \citep{Tarricq2022}.
\\
The central problem addressed in this review is that these detections are not directly interchangeable. A ``member'' can mean a star passing a kinematic cut, a star with a membership probability above a chosen threshold, a point assigned to a density cluster, or a star whose orbit resembles that of an N-body model. Different definitions can yield different tail lengths and even qualitatively distinct morphologies. For example, a method designed to identify non-Gaussian low-density structures can retain more distant tail stars than a model that assumes a compact Gaussian distribution, while a purely kinematic selection can include co-moving field stars that are not dynamically associated with the cluster.
\\
This methodological dependence has become particularly important after Gaia DR3. \citet{Kos2024} applied a probabilistic model of cluster dissolution to 476 open clusters and found high-probability tail members for most systems in a selected sample, with tails extending beyond a kiloparsec in some cases. \citet{Risbud2025} used a convergent-point approach for 21 nearby clusters and reported tidal tails in 19 of them, with typical spans of 20--200 pc. \citet{Jadhav2025} subsequently examined the reliability of observed catalogues using simulations and emphasized that observational catalogues can be incomplete and method-dependent. These results substantially change the context in which earlier individual-cluster studies should be interpreted.
\\
As a review paper, the purpose of this work is therefore not to compile a simple list of tail detections. Instead, we ask four related questions: (1) what physical and observational criteria distinguish a tidal tail from a generic extended cluster population; (2) how do the main membership and tail-detection methods operate and what assumptions do they impose; (3) where do published studies of the same cluster agree or disagree quantitatively; and (4) what combination of observables is required for a robust physical interpretation? By answering these questions, we aim to provide a framework for comparing the heterogeneous Gaia-era literature.

\section{Physical framework and terminology}

\subsection{Tidal radius and escape}

The classical tidal or Jacobi radius provides a useful scale for discussing escape. In an axisymmetric Galactic potential, a simplified estimate near the solar neighbourhood can be written in terms of the cluster mass and the local angular frequency and Oort constants. For a cluster of mass $M$, the characteristic escape radius is of order
\begin{equation}
r_{\rm J} \sim \left[\frac{GM}{4A(A-B)}\right]^{1/3},
\end{equation}
where $A$ and $B$ are the Oort constants \citep{RoserSchilbach2019}. The exact value depends on the Galactic potential and the cluster orbit, and $r_{\rm J}$ should not be interpreted as an observational boundary separating ``members'' from ``non-members'' with absolute certainty.
\\
Stars can cross the Jacobi surface through the vicinity of the inner and outer Lagrange points. Their subsequent motion is influenced by the cluster, the Galactic tidal field, and the Coriolis force in the rotating frame. Numerical calculations show that the resulting structures can remain coherent over hundreds of parsecs, depending on cluster age, orbit, mass-loss history, and the local Galactic environment \citep{Kupper2010}.

\subsection{What is, and is not, a tidal tail?}

The literature contains several related but distinct concepts. A corona or halo can refer to an extended stellar population surrounding the cluster without requiring that the stars are unbound. An extra-tidal population simply denotes stars beyond a chosen structural radius. A co-moving group is a kinematic description and does not by itself establish a common dynamical origin. A tidal tail is a stronger physical interpretation: it is an extended population plausibly produced by the escape and subsequent orbital evolution of cluster stars.
\\
This distinction is central to the comparison performed here. A spatially elongated distribution outside a fitted King or Plummer radius is not, by itself, sufficient evidence for a tidal tail. Conversely, a tail can be very difficult to recognize as a density enhancement if it is viewed nearly along the line of sight or if phase mixing has substantially reduced its contrast. Studies based on two-dimensional projected positions, therefore, probe a different quantity from studies using five- or six-dimensional phase space \citep{Tarricq2022}.

\section{Literature search and comparison framework}

\subsection{Scope and review strategy}

The present review is designed as a methodological and comparative synthesis rather than as a catalogue of detections. We searched the literature up to September 2026 using combinations of the terms ``open cluster'', ``tidal tail'', ``extra-tidal'', ``corona'', ``extended structure'', ``Gaia'', ``membership'', ``convergent point'', ``DBSCAN'', ``HDBSCAN'', ``Gaussian mixture'', ``orbit'', and ``N-body''. We used the recent systematic compilations of \citet{Jadhav2025} as starting points and then checked the individual studies cited by those works, together with newer 2025--2026 publications. This is important because the literature has expanded rapidly, and several recent papers now contain information that was unavailable when earlier reviews were written.
\\
The comparison is deliberately hierarchical. First, we distinguish an extended population from a dynamically established tidal tail. Second, we compare the data and membership method used to obtain each population. Third, we record quantitative observables such as the number of selected stars, tail extent, tidal radius, and contamination whenever the source paper reports them. Finally, we assess whether the proposed interpretation is supported by independent photometric, spectroscopic, chemical, or dynamical evidence. The resulting table should therefore be read as a structured comparison of published measurements, not as a merged catalogue.
\\
A particularly important benchmark is \citet{Jadhav2025}, who compiled 122 published catalogues corresponding to 58 unique clusters within 500 pc and graded them using photometric, morphological, and dynamical tests. They found 15 gold, 55 silver, and 51 bronze catalogues. We use this study as an external reliability benchmark and focus our own synthesis on the methodological reasons for disagreement between catalogues and on the physical meaning of the different evidence levels. This distinction is essential because a review that simply repeats a large catalogue of detections would add little beyond the existing literature.
\\
We organize the literature according to both chronology and methodology. The comparison includes theoretical studies of escape and tail formation, pre-Gaia observational work, Gaia-era individual-cluster analyses, systematic Gaia catalogues, and recent methodological studies. Particular attention is given to papers that report (i) an explicit membership method, (ii) a quantitative tail extent or member count, (iii) a comparison with a dynamical model, or (iv) a systematic sample of clusters.
\\
For each study, we distinguish the following quantities whenever they are available: Gaia data release; dimensionality of the data; membership algorithm; membership threshold or selection criterion; number of core and tail candidates; reported tail length; use of photometric, spectroscopic, or chemical validation; and use of orbit or N-body modeling. We do not combine member counts from different studies into a single ``best'' value because the samples are defined differently.

\section{Overview of the observational literature}

\subsection{The pre-Gaia period}

The theoretical expectation of tidal tails predates their direct detection by many decades. 
\citet {Bok1934} and \citet{Spitzer1940} provided early descriptions of cluster dynamics in an external tidal field. Later analytical and N-body work established the importance of escape through the Lagrange points and the development of epicyclic overdensities \citep{Kupper2008}.
\\
Before Gaia, open-cluster tail searches were restricted mainly to nearby clusters and depended on photometry and heterogeneous proper-motion catalogues. The main difficulty was the absence of a sufficiently precise and homogeneous phase-space reference population. A low-density halo could be interpreted as a genuine extended cluster population, field contamination, or residual structure inherited from star formation. Consequently, pre-Gaia detections did not provide a homogeneous census.

\subsection{The Gaia DR2 transition}

Gaia DR2 represented a major change because parallaxes and proper motions could be combined with broad-band photometry for large numbers of stars. \citet{Roser2019} applied a modified convergent-point method to the Hyades and identified a leading tail extending to about 170 pc from the cluster centre and a trailing tail to about 70 pc. Their selected sample contained 292 leading-tail candidates and 237 trailing-tail candidates, with estimated contamination of 36 and 32 stars, respectively. \citet{MeingastAlves2019} independently recovered a highly elongated Hyades structure in six-dimensional phase space.
\\
The Coma Berenices cluster provided another important test. \citet{Tang2019} used the StarGO neural-network clustering method in five-dimensional space and reported leading and trailing structures of about 50 pc each. Within the tidal radius, they identified 77 member candidates, while about 120 additional candidates were assigned to the tidal structures. The large fraction of stars outside the core illustrated both the physical importance of extra-tidal populations and the sensitivity of inferred membership to the selection method.
\\
\citet{Gao2020a,Gao2020b} demonstrated a different strategy by combining principal-component analysis with Gaussian mixture models. For M\,67, 1618 stars were identified as closely related to the cluster, of which 85 were considered likely extra-tidal; the reported tails reached at least about 39 pc in projection. For NGC 2506, the same general strategy identified a larger population of candidate extra-tidal stars. These studies illustrate the strength of probabilistic approaches for producing reproducible membership probabilities, but also the danger that a compact Gaussian representation may under-represent highly non-Gaussian tails.

\subsection{EDR3 and DR3: from individual systems to systematic searches}

\citet{Meingast2021} developed a bulk-velocity-informed membership method for ten nearby young clusters and identified very extended stellar coronae, often exceeding 100 pc. Importantly, they emphasized that these populations can reflect differential Galactic rotation and early cluster evolution; they should therefore not automatically be equated with mature tidal tails. Their results helped establish that extended structures are not restricted to the Hyades. \citet{Tarricq2022} subsequently analysed 389 local clusters and used HDBSCAN to identify elongated structures in a homogeneous sample. Their work included 71 nearby clusters with identified tidal tails and found previously reported structures around systems, including Coma Berenices, Blanco 1, NGC 752, NGC 7092, and NGC 2516. Importantly, their two-dimensional projected analysis could miss tails oriented close to the line of sight, demonstrating that non-detection is not equivalent to physical absence.
\\
The all-sky Gaia DR3 catalogue of \citet{HuntReffert2023} used HDBSCAN to recover 7167 clusters, including 2387 candidate new objects, and produced membership lists that often extend beyond conventional cluster radii. Their catalogue illustrates another important point: a membership catalogue optimized for finding clusters is not necessarily optimized for identifying tidal tails. Nevertheless, their non-spherical treatment of cluster structure enables recovery of extended features that spherical or compact-profile models may suppress.
\\
\citet{Kos2024} made the first particularly broad probabilistic tail search based on a simulated cluster-dissolution model. The study considered 476 open clusters with $G<17.5$ and, for the selected sample, ages above 100 Myr and distances of up to 3 kpc. High-probability tail members were found for most systems, with some reconstructed tails extending more than 1 kpc. This result demonstrates that the physical size of a tidal structure can be much larger than the few tens of parsecs typically measured in nearby individual-cluster studies, but it also shows why tail catalogues must be interpreted together with their selection functions and membership probabilities.
\\
\citet{Risbud2025} used Gaia DR3 and a convergent-point method to analyse 21 clusters within 400 pc, older than 100 Myr, and with more than 100 members. They reported tails in 19 clusters, with typical spans of 20--200 pc and 30--700 stars in the region comprising the tidal radius and tails. Their self-compact convergent-point (SCCP) variant was designed to improve tail tracing. The same study also identified cases in which tail orientations and velocity properties challenge simple expectations, emphasizing the need for dynamical validation.
\\
\citet{Jadhav2025} reviewed the reliability of nearby-tail catalogues using simulations. Their analysis is especially relevant to the interpretation of the high detection rate reported in \citep{Risbud2025}: the observed number of tails cannot be interpreted independently of completeness, contamination, cluster orientation, and the adopted method's ability to recover diffuse structures.
\\
Recent work has continued to expand the sample. \citet{Sharma2025} analysed five clusters (BH 164, Alessi 2, NGC 2281, NGC 2354, and M\,67) with Gaia DR3 using CMD matching, PCA, DBSCAN, and additional statistical diagnostics. They reported tails of roughly 40--100 pc containing about 100--200 stars. These results reinforce the conclusion that extended structures are widespread and show that the detailed tail morphology depends on the analysis pipeline.

\section{Methodological foundations}

\subsection{Convergent-point and co-moving selection}

The convergent-point method is particularly intuitive for nearby clusters with a measurable bulk motion. A star moving with the cluster should have a proper-motion component directed toward the cluster's convergent point. The observed proper-motion vector can therefore be decomposed into components parallel and perpendicular to the expected direction. Candidate members are selected when the observed and predicted tangential velocities agree within a specified tolerance \citep{Roser2019}.
\\
Its main strength is that it directly exploits the kinematic coherence of escaping stars and does not require a parametric spatial density model. Its principal weakness is that field stars can share similar proper motions, especially over limited sky areas. The method also requires careful treatment of distance and measurement uncertainties. \citet{Risbud2025} introduced the SCCP approach to improve compactness and tail tracing, illustrating how even within the same broad methodological family, the selection function can change the recovered morphology.

\subsection{Density-based clustering: DBSCAN and HDBSCAN}

DBSCAN identifies density-connected groups using a neighbourhood scale and a minimum number of neighbours \citep{Ester1996}. Unlike a Gaussian model, it does not require a spherical or elliptical cluster profile and can therefore follow elongated structures. However, the recovered membership depends on the adopted density scale and minimum cluster size. A real tail whose density decreases continuously can fragment into several components or disappear below the adopted threshold.
\\
HDBSCAN extends the density-based idea by constructing a hierarchy of density-connected structures and extracting clusters across a range of density levels \citep{Campello2015}. This is attractive for Galactic cluster fields because the density of the field and cluster tails can vary strongly with position. HDBSCAN has been used in large Gaia cluster censuses and can recover non-spherical structures \citep{HuntReffert2023}. Its output, however, remains an algorithmic description of the density structure. A low-density cluster can be linked to unrelated field structures if the phase-space metric and feature scaling are not physically appropriate.

\subsection{Gaussian mixture models}

A GMM represents the observed distribution as a weighted sum of Gaussian components. In cluster membership studies, one or more components can represent the cluster, and the remaining components the field. Membership is expressed probabilistically rather than through a hard geometric boundary. ML-MOC, for example, combines k-nearest neighbours and GMMs to derive membership probabilities from Gaia astrometry \citep{Bhattacharya2022}.
\\
The major advantage is interpretability and probabilistic membership. The major limitation for tidal tails is structural: a Gaussian component naturally emphasizes the high-density core and can assign low probability to a long, curved, or multi-modal tail. PCA+GMM analyses can partially alleviate dimensionality issues, but the basic Gaussian assumption still holds. Therefore, a compact result from a GMM study should not automatically be interpreted as evidence that the cluster lacks extended tidal debris.

\subsection{k-nearest neighbours and supervised or hybrid machine learning}

kNN methods classify stars according to their similarity to labeled or high-confidence members. They are flexible and can operate in multi-dimensional astrometric space, but their output depends on the training set and the local data density. ML-MOC demonstrated that kNN+GMM can provide membership probabilities down to approximately $G=20$ for a range of clusters \citep{Agarwal2021}. Such methods are valuable for membership refinement, but the training or reference population can implicitly encode a compact-cluster assumption.
\\
Recent hybrid approaches combine clustering, PCA, CMD information, and supervised learning. \citet{Sharma2025} illustrate this strategy by combining DBSCAN and PCA with photometric and kinematic diagnostics. These approaches are promising because a physically meaningful tail should be simultaneously coherent in several observable spaces.

\subsection{Orbit integration and N-body matching}

\begin{figure*}
\centering
\begin{tikzpicture}[scale=0.86]
\node[draw,rounded corners,align=center,minimum width=3.4cm,minimum height=1cm] (data) at (0,0)
{Gaia astrometry\\parallax + proper motion\\+ photometry};
\node[draw,rounded corners,align=center,minimum width=3.4cm,minimum height=1cm] (kin) at (-5,-2.2)
{Kinematic selection\\CP / co-moving};
\node[draw,rounded corners,align=center,minimum width=3.4cm,minimum height=1cm] (dens) at (0,-2.2)
{Density clustering\\DBSCAN / HDBSCAN};
\node[draw,rounded corners,align=center,minimum width=3.4cm,minimum height=1cm] (prob) at (5,-2.2)
{Probabilistic models\\GMM / kNN / hybrid ML};
\node[draw,rounded corners,align=center,minimum width=3.4cm,minimum height=1cm] (phys) at (0,-4.5)
{Physical validation\\CMD + RV + chemistry\\+ orbit / N-body};
\node[draw,rounded corners,align=center,minimum width=4.2cm,minimum height=1cm] (tail) at (0,-6.5)
{Robust tidal-tail interpretation};
\draw[-{Latex[length=3mm]}] (data) -- (kin);
\draw[-{Latex[length=3mm]}] (data) -- (dens);
\draw[-{Latex[length=3mm]}] (data) -- (prob);
\draw[-{Latex[length=3mm]}] (kin) -- (phys);
\draw[-{Latex[length=3mm]}] (dens) -- (phys);
\draw[-{Latex[length=3mm]}] (prob) -- (phys);
\draw[-{Latex[length=3mm]}] (phys) -- (tail);
\end{tikzpicture}
\caption{Gaia astrometry provides the initial phase-space information, while kinematic, density-based, and probabilistic methods define candidate populations in different ways. A robust physical interpretation should then be tested using independent photometric, radial-velocity, chemical, and dynamical information. The diagram emphasizes that membership selection and physical validation are distinct stages of the analysis.}
\label{fig:workflow}
\end{figure*}
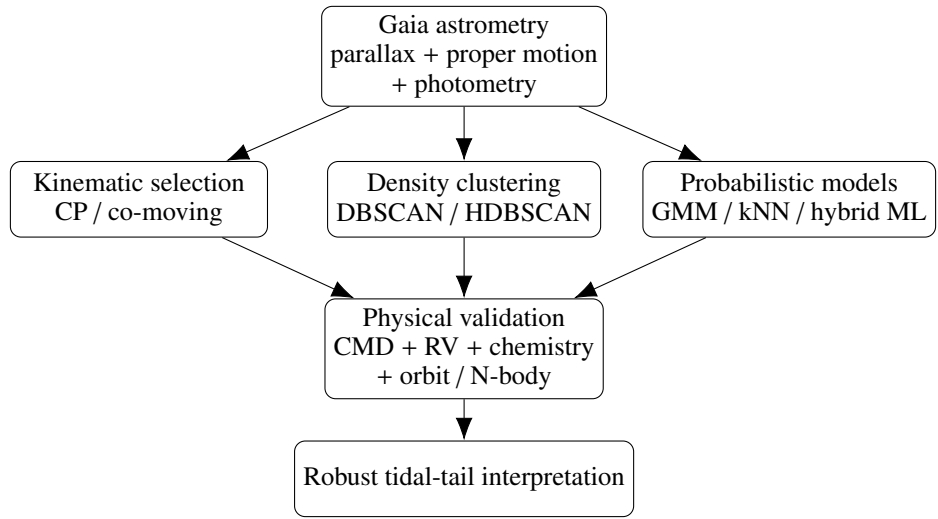

Orbit-based methods impose the strongest physical prior. Candidate stars are integrated in an assumed Galactic potential and compared with the cluster orbit or with an N-body model of cluster dissolution. This can distinguish a tidal tail from an unrelated co-moving group when the six-dimensional information is sufficiently precise. N-body modeling also predicts escape directions, tail curvature, density variations, and the expected relation between cluster properties and tail morphology \citep{Kupper2010}.
\\
The drawback is model dependence. The result depends on the assumed Galactic potential, cluster mass, initial conditions, binary population, and treatment of the external environment. A model can therefore reject genuine stars if its assumptions are inaccurate. Conversely, fitting a model to the data can make the resulting structure appear more physically secure than the observational constraints justify. The best use of orbit/N-body methods is therefore as a validation layer rather than as a replacement for empirical membership evidence.
\\
The recommended hierarchy for tidal-tail identification is shown in Figure \ref{fig:workflow}.

\section{Comparison of the literature}

\subsection{Hyades: benchmark but not method-independent}

The Hyades provide the clearest case of extended tidal structure. \citet{Roser2019} reported a leading tail to about 170 pc and a trailing tail to about 70 pc, with 292 and 237 candidate stars, respectively. \citet{MeingastAlves2019} independently identified an elongated structure using six-dimensional information. Later work extended the detected structures and investigated their substructure and asymmetry \citep{Risbud2025}.
\\
The important lesson is that ``confirmed'' does not mean ``uniquely measured''. Tail length varies because different studies use different search volumes, probability thresholds, velocity tolerances, and data releases. Nevertheless, the existence of extended Hyades debris is unusually robust because independent methods recover a coherent leading/trailing configuration and because the observed structure can be compared with dynamical models.

\subsection{Praesepe: extended population with detailed structural modeling}

Praesepe is another mature benchmark. \citet{RoserSchilbach2019} analysed 1393 stars associated with the cluster and tails. They reported both leading and trailing tails extending to about 165 pc from the cluster centre, with 1393 stars in the cluster-plus-tail sample. Their tidal radius was 10.77 pc, with 725 stars inside that radius and a tidal mass of $483\,M_\odot$; they estimated 389 stars beyond twice the tidal radius and a field contamination of about 3.6--7.2\%. The study also demonstrated mass segregation and identified white dwarfs associated with the cluster and its extended population.
\\
The interpretation of Praesepe has since expanded from empirical membership to dynamical modeling. Recent N-body work tailored to the cluster investigates its mass profile, stellar mass function, mass segregation, and tidal-tail kinematics \citep{Weis2025}. This makes Praesepe an important example of the transition from ``tail detection'' to ``tail modeling. At the same time, its results illustrate why membership counts from different methods cannot be compared without documenting the selection function.

\subsection{Coma Berenices: a strong extended structure but sensitive to definition}

\citet{Tang2019} identified about 115 $M_\odot$ in the cluster and approximately 120 additional member candidates in the tidal structures, with leading and trailing extensions of about 50 pc each. Their StarGO analysis used five-dimensional phase-space information. Later systematic analyses recovered extended populations but also demonstrated that the measured morphology depends on the dimensionality and membership criteria \citep{Risbud2025}.
\\
Coma Berenices is therefore better described as a robust, extended, co-moving population with a tidal interpretation strongly supported by the dynamics, rather than as a system in which every study recovers an identical tail boundary.

\subsection{Blanco 1, Pleiades, and NGC 2516: young systems and the primordial-structure problem}

Blanco 1 provides a useful reference for young clusters. \citet{Li2021} described two tails of roughly 50 pc and used the cluster as a reference for assessing early-stage tidal disruption in the Pleiades and NGC 2516. For the latter two systems, the majority of the mass was still estimated to lie inside the tidal radius, consistent with an early disruption stage.
\\
The interpretation is important because elongated structures in young clusters do not automatically imply long-term tidal evaporation. Residual primordial substructure, gas-expulsion-driven expansion, and the imprint of the birth environment can survive for a significant fraction of the early dynamical history. \citet{Tarricq2022} likewise found that the morphology of young clusters can retain signatures of their formation history, whereas older clusters more readily show structures attributable to tidal stripping.

\subsection{M\,67, NGC 2506, NGC 752, and IC 4756: complementary methods}

M\,67 illustrates the use of PCA+GMM. \citet{Gao2020a} identified 85 likely extra-tidal stars among 1618 stars closely related to the cluster, with tails extending to about 39 pc in projection. NGC 752 provides a contrasting example in which a broader search region and a convergent-point-plus-DBSCAN analysis substantially increased the inferred tail extent. \citet{Boffin2022} reported tails extending more than 260 pc tip-to-tip, about four times the previously claimed extent, and used N-body calculations as a guide.
\\
IC 4756 demonstrates the value of combining three-dimensional positions, tangential velocities, and photometric validation. \citet{Ye2021} identified 644 members and reported two S-shaped tails extending to about 180 pc, with a tidal radius of about 12 pc. These studies show that the same physical question can be approached through fundamentally different selection functions.

\subsection{The expanding sample of nearby clusters}

Systematic catalogues have broadened the sample beyond the classic nearby examples. \citet{Tarricq2022} characterized extended structures in a large local sample. \citet{Kos2024} found high-probability tail populations in most of the 476 analysed clusters and reported tail scales exceeding 1 kpc in some systems. \citet{Risbud2025} found tails in 19 of 21 nearby clusters satisfying their selection criteria. \citet{Sharma2025} reported tails of 40--100 pc in five additional systems.
\\
The most comprehensive reliability analysis to date is Jadhav et al. \citep{Jadhav2025}. Their compilation contains 122 tail catalogues for 58 unique clusters within 500 pc. Rather than assuming that every published catalogue is equally reliable, they graded the catalogues using a combination of photometric, morphological, and dynamical diagnostics. Only 15 catalogues reached their highest ``gold'' grade, 55 were classified as ``silver'', and 51 as ``bronze''. This result is particularly important for interpreting the high detection rates obtained by recent Gaia searches: a high fraction of detections does not imply a high fraction of equally secure physical tails.
\\
The literature has continued to grow in 2026. \citet{Zhao2026} analysed 65 nearby clusters within 500 pc and older than 50 Myr using a homogeneous Gaia DR3 pipeline that combines nearest-neighbour/DBSCAN candidate selection with a subsequent Forest Fire Clustering membership stage. Their study links tail properties with mass segregation and illustrates the value of analysing cluster structure and tails together. In parallel, \citet{Parul2026} investigated how the Galactic bar can deflect tidal tails and showed that tail orientation itself can depend on the non-axisymmetric Galactic potential. These studies broaden the interpretation of tidal tails from simple dissolution tracers toward probes of Galactic dynamics.
\\
Systematic catalogues have broadened the sample beyond the classic nearby examples. \citet{Tarricq2022} characterized extended structures in a large local sample. \citet{Kos2024} found high-probability tail populations in most of the 476 analysed clusters and reported tail scales exceeding 1 kpc in some systems. \citet{Risbud2025} found tails in 19 of 21 nearby clusters satisfying their selection criteria. \citet{Sharma2025} reported tails of 40--100 pc in five additional systems.
\\
These high detection rates should not be interpreted as evidence that almost all open clusters possess equally well-defined tidal tails. Instead, they demonstrate that extended cluster-associated populations are common. Whether an extended population is a classical two-sided tidal tail, a dissolving association, a co-moving stream, or a mixture of these components remains to be validated.

\begin{table*}
\caption{Representative quantitative results from Gaia-era tidal-tail studies. Numbers are reproduced as reported in the cited studies and should not be treated as homogeneous measurements, as the membership definitions and search regions differ.}
\label{tab:quantitative}
\resizebox{\textwidth}{!}{
\centering
\begin{tabular}{l l l l l l}
\hline
Cluster & Reference & Gaia & Method & Tail extent & Membership information\\
\hline
Hyades & \citet{Roser2019} & DR2 & CP & $+170/-70$ pc & 292 / 237 candidates\\
Hyades & \citet{MeingastAlves2019} & DR2 & 6D selection & several hundred pc & study-specific\\
Praesepe & \citet{RoserSchilbach2019} & DR2 & astrometry + photometry & extended tails & 1393 associated stars; 725 inside $r_t$\\
Coma Ber & \citet{Tang2019} & DR2 & StarGO & $\sim$50 pc each side & 77 core; $\sim$120 tail candidates\\
M\,67 & \citet{Gao2020a} & DR2 & PCA + GMM & $\geq$39 pc projected & 1618 related; 85 extra-tidal\\
Blanco 1 & \citet{Li2021} & EDR3 & kinematic & $\sim$50 pc each side & 63\% mass inside $r_t$\\
IC 4756 & \citet{Ye2021} & EDR3 & 5D + Bayesian distance & $\sim$180 pc & 644 members\\
NGC 752 & \citet{Boffin2022} & EDR3 & CP + DBSCAN & $>260$ pc tip-to-tip & about twice previous tail count\\
Nearby OCs & \citet{Tarricq2022} & EDR3 & HDBSCAN & sample dependent & 71 nearby OCs with tails\\
476 OCs & \citet{Kos2024} & DR3 & probabilistic model & $>$1 kpc in some cases & catalogue of high-probability tail stars\\
21 nearby OCs & \citet{Risbud2025} & DR3 & CP/SCCP & 20--200 pc typical span & 30--700 stars in $r_t+$tail region\\
5 OCs & \citet{Sharma2025} & DR3 & PCA + DBSCAN + statistics & 40--100 pc & $\sim$100--200 tail stars\\
65 nearby OCs & \citet{Zhao2026} & DR3 & DBSCAN + FFC & sample dependent & homogeneous 65-OC analysis\\
\hline
\end{tabular}}
\end{table*}

\subsection{What the quantitative comparison actually shows}

Three conclusions emerge from Table~\ref{tab:quantitative}. First, tail extent is strongly dependent on both method and search volume. A study restricted to a few tens of parsecs cannot recover a 200 pc tail even if that tail exists. Second, the number of ``tail members'' is not a physical invariant. A probability threshold, density cut, or velocity tolerance can substantially change the count. Third, a large tail catalogue and a high-confidence individual detection answer different questions.
\\
The appropriate unit of comparison is therefore not simply the number of stars or the tail length. A more informative comparison includes the completeness and contamination of the selection, the dimensionality of the data, the location of the tail relative to the Jacobi radius, and the independent evidence that the selected stars share the cluster's age and chemical composition. This is particularly important when comparing density-based catalogues with convergent-point or dynamical-model catalogues.

\section{Why published detections disagree}

\subsection{Different definitions of membership}

The most fundamental difference is the meaning of ``member''. A hard proper-motion cut, a 0.5 membership probability, an HDBSCAN cluster label, and a star passing an N-body phase-space filter are not equivalent criteria. In a dense core, these definitions often converge. In a diffuse tail, they can diverge dramatically.

\subsection{Dimensionality and projection}

A tail is a three-dimensional spatial structure embedded in a six-dimensional phase-space distribution. A two-dimensional sky map can therefore hide a real tail if it is aligned close to the line of sight, while a projected elongation can be produced by unrelated stars at different distances. \citet{Tarricq2022} explicitly demonstrated that a 2D analysis can fail to recover tails that are strongly elongated along the line of sight.

\subsection{Field contamination}

The field population is not spatially or kinematically uniform. Moving groups, dissolved clusters, spiral-arm populations, and Galactic shear can produce coherent structures that resemble cluster debris. COIN-Gaia 13 provides an instructive case: an elongated structure of about 270 pc was identified, but its spatial and kinematic properties were argued to be more naturally explained by differential Galactic rotation than by a conventional tidal tail \citep{Bai2022}. This illustrates why ``extended'' and ``tidal'' should not be treated as synonyms.

\subsection{Data-release and completeness effects}

Gaia DR2, EDR3, and DR3 differ in precision, completeness, and radial-velocity content. The improvement is particularly important for faint tail stars, but completeness is also strongly dependent on distance, extinction, crowding, and stellar magnitude. \citet{HuntReffert2023} show that Gaia DR3 can substantially enlarge membership lists for nearby clusters, while the gain is much smaller for distant clusters where astrometric errors dominate.

\subsection{Model dependence}

Probabilistic and orbit-based methods introduce explicit assumptions. A GMM assumes that the selected distribution can be represented adequately by Gaussian components; an N-body model assumes a cluster mass, initial state, Galactic potential, and evolutionary history. These assumptions are useful because they add physical information, but they also make the result conditional on the model. The most reliable conclusions are therefore those that survive comparison across methods with different assumptions.

\section{Cross-study comparison: what can be compared quantitatively?}

A major difficulty in the literature is that apparently similar quantities are often defined differently. We therefore recommend reporting at least six quantities for every future tail catalogue: (1) the search radius or volume; (2) the number of core members and candidate tail members; (3) the adopted membership probability or kinematic tolerance; (4) the projected and three-dimensional tail extent; (5) an estimate of contamination or completeness; and (6) whether independent photometric, spectroscopic, chemical, or dynamical tests were applied.
\\
The distinction between projected and orbital tail length is especially important. A curved tail can have a substantially larger arc length than its linear extent measured by a principal component or straight-line projection. Similarly, a ``tip-to-tip'' span is not directly comparable to a one-sided distance from the cluster centre. We therefore retain the definitions used by the source papers rather than converting all measurements to a common quantity without the underlying catalogues.
\\
The recent literature also demonstrates that the number of tail members is not a stable physical observable. \citet{Roser2019} selected 292 leading- and 237 trailing-tail candidates for the Hyades, whereas \citet{Tang2019} identified about 120 candidates in the Coma Berenices tidal structures. \citet{Risbud2025} report 30--700 stars in the combined tidal-radius and tail regions across their nearby-cluster sample, while \citet{Sharma2025} find about 100--200 tail stars in five Gaia DR3 clusters. These values cannot be ranked without considering the depth, spatial search region, probability threshold, and completeness of each catalogue.
\\
\citet{Jadhav2025} provides an especially useful independent test because they explicitly grade catalogues rather than treating all detections as equivalent. Their result that only 15 of 122 compiled catalogues satisfy their highest-grade criteria is a quantitative indication of the current uncertainty. The appropriate conclusion is not that the remaining catalogues are false, but that the evidence supporting their physical interpretation is weaker or incomplete.

\subsection{Completeness, contamination, and non-detections}

A non-detection is not symmetric with a detection. A method may fail to recover a tail because the tail is aligned close to the line of sight, because its surface density falls below the field background, because the stars are too faint for reliable astrometry, or because the adopted feature space does not preserve the relevant phase-space correlations. Conversely, a detection may be contaminated by co-moving field stars or structures produced by Galactic shear. Consequently, catalogue completeness and contamination should be estimated with injected or simulated clusters whenever possible. The simulation-based analysis of \citet{Jadhav2025} demonstrates why such calibration is essential.

\section{A proposed hierarchy of evidence}

We propose a practical classification for future studies.

\begin{enumerate}
\item \textbf{Level 0 -- spatial extension:} an overdensity or elongation is detected beyond a conventional cluster radius.
\item \textbf{Level 1 -- astrometric coherence:} the extended stars share parallax and proper-motion properties with the cluster.
\item \textbf{Level 2 -- photometric consistency:} the extended stars occupy the cluster sequence in a colour--magnitude diagram after accounting for extinction and binaries.
\item \textbf{Level 3 -- dynamical consistency:} the stars are consistent with escape trajectories in an orbit or N-body model.
\item \textbf{Level 4 -- independent spectroscopic/chemical confirmation:} radial velocities and/or chemical abundances support a common origin.
\end{enumerate}

This hierarchy avoids treating every extended population as equally secure. It also allows different studies to be compared even when they do not use the same algorithm. A catalogue can report Level 1 candidates without claiming Level 4 confirmation. 
\\
Appendix A represents the literature comparison tables.

\section{Future prospects}

Gaia DR4 is currently expected in December 2026 and will be based on 66 months of observations; the Gaia mission itself ceased science observations in January 2025. The longer time baseline and expanded data products should improve proper motions and parallaxes and should be particularly valuable for low-contrast structures \citep{GaiaMission2026}. The main scientific gain will not simply be more stars but better velocity precision and a larger fraction of stars with sufficiently useful astrometry for dynamical tests.
\\
Large spectroscopic surveys such as 4MOST and WEAVE can provide radial velocities and chemical abundances for stars in cluster outskirts. Chemical homogeneity is especially valuable for distinguishing true cluster debris from unrelated co-moving field populations. For older and more distant clusters, spectroscopy may be the only practical way to reduce contamination to a level where tail morphology becomes physically interpretable.
\\
Deep photometric surveys, including Rubin Observatory/LSST, will extend the detectable stellar mass range beyond Gaia's effective completeness limits. Combining deep photometry with Gaia astrometry can therefore reveal the low-mass stars that are expected to dominate the number counts in tidal debris.
\\
The next methodological step should be a benchmarked multi-method analysis. The same Gaia sample should be processed with at least one density-based, one probabilistic, one kinematic, and one dynamical method. The resulting catalogues should be cross-matched star-by-star. Such an experiment would directly quantify method-specific completeness and contamination and would provide a more meaningful uncertainty estimate for tail length than the spread across unrelated literature studies.
\\
A second priority is simulation-based calibration. \citet{Jadhav2025} demonstrates the importance of understanding observational catalogue reliability. Synthetic clusters with known tail membership can be injected into realistic Gaia-like fields and recovered with different algorithms. This would allow the community to estimate the probability that a reported tail is real as a function of distance, age, orientation, surface density, and Galactic environment.
\\
Finally, the field should move from static morphology toward time-dependent modeling. Coupling Gaia-constrained initial conditions with N-body simulations can test whether observed overdensities are generated by steady escape, epicyclic motion, molecular cloud encounters, spiral armthethe perturbations, or observational selection. This is the route by which tidal tails can become quantitative probes of both cluster evolution and the Galactic potential.

\section{Conclusions}

The Gaia era has transformed the study of tidal tails of Galactic open clusters. The observational problem has progressed from isolated searches for a few nearby systems to systematic catalogues containing tens or hundreds of candidate extended structures. The physical interpretation, however, has become more—not less—subtle.
\\
The Hyades remains the benchmark for a robust tidal-tail detection because independent analyses recover coherent leading and trailing structures and because the observations can be compared with dynamical models. Praesepe and Coma Berenices also provide strong evidence for extended cluster-associated populations, while M\,67, NGC 752, IC 4756, Blanco 1, NGC 2516, and many other clusters demonstrate the diversity of tail morphology and the importance of the adopted selection method.
\\
The principal result of this review is that a tidal-tail catalogue should not be interpreted as a homogeneous list of binary detections. Published tail lengths, member counts, and morphologies depend on the search region, Gaia data release, data dimensionality, membership definition, probability threshold, clustering algorithm, and dynamical assumptions. Recent homogeneous studies demonstrate that extended structures are widespread, but they do not imply that every reported structure has the same physical origin or the same level of observational significance.
\\
A useful future standard is therefore to separate detection from interpretation. Astrometric coherence should establish candidate membership; photometric consistency should test the stellar population; radial velocities and chemical abundances should provide independent confirmation; and orbit or N-body calculations should test physical plausibility. The strongest tidal-tail detections are those supported by several independent layers of evidence.
\\
The field is now ready for a benchmark analysis in which the same Gaia sample is processed with multiple membership methods and calibrated against simulated clusters. Such a comparison would directly quantify algorithmic bias, contamination, and completeness. Combined with Gaia, deep photometry, and large spectroscopic surveys, this approach should turn tidal tails from visually compelling but method-dependent structures into quantitatively calibrated tracers of cluster dissolution and Galactic dynamics.

\begin{acknowledgements}
This work was supported by the grant GAčR 23-07605S.
\end{acknowledgements}

\bibliographystyle{aa}
\bibliography{paper}

@article{Agarwal2021,
  author = {Agarwal, Manan and Rao, Khushboo K. and Vaidya, Kaushar and Bhattacharya, Souradeep},
  title = {ML-MOC: Machine Learning (kNN and GMM) based Membership Determination for Open Clusters},
  journal = {\mnras}, volume = {502}, pages = {2582--2602}, year = {2021}, doi = {10.1093/mnras/stab095}}

@article{Bai2022,
  author = {Bai, Leya and Zhong, Jing and Chen, Li and Li, Jing and Hou, Jinliang},
  title = {Discovery of Extended Structure Around Open Cluster COIN-Gaia 13 Based on Gaia EDR3},
  journal = {Research in Astronomy and Astrophysics}, volume = {22}, pages = {055007}, year = {2022}, doi = {10.1088/1674-4527/ac60d2}}

@article{Bhattacharya2022,
  author = {Bhattacharya, Souradeep and Rao, Khushboo K. and Agarwal, Manan and Balan, Shanmugha and Vaidya, Kaushar},
  title = {A Gaia EDR3 search for tidal tails in disintegrating open clusters},
  journal = {\mnras}, volume = {517}, pages = {3525--3542}, year = {2022}}

@article{Boffin2022,
  author = {Boffin, Henri M. J. and Jeřábková, Tereza and Beccari, Giacomo and Wang, Long},
  title = {A tale of caution: the tails of NGC 752 are much longer than claimed},
  journal = {\mnras}, volume = {514}, pages = {3279--3288}, year = {2022}}

@article{Bok1934,
  author = {Bok, Bart J.},
  title = {The distribution of stars in space},
  journal = {Harvard College Observatory Circular}, volume = {384}, pages = {1--10}, year = {1934}}

@article{Campello2015,
  author = {Campello, Ricardo J. G. B. and Moulavi, Davoud and Zimek, Arthur and Sander, Jörg},
  title = {Hierarchical Density Estimates for Data Clustering, Visualization, and Outlier Detection},
  journal = {ACM Transactions on Knowledge Discovery from Data}, volume = {10}, pages = {5}, year = {2015}, doi = {10.1145/2733381}}

@article{Ernst2011,
  author = {Ernst, Andreas and Just, Andreas and Berczik, Peter and Olczak, Christopher},
  title = {The tidal tails of star clusters},
  journal = {A\&A}, volume = {536}, pages = {A64}, year = {2011}}

@inproceedings{Ester1996,
  author = {Ester, Martin and Kriegel, Hans-Peter and Sander, Jörg and Xu, Xiaowei},
  title = {A Density-Based Algorithm for Discovering Clusters in Large Spatial Databases with Noise},
  booktitle = {Proceedings of the Second International Conference on Knowledge Discovery and Data Mining},
  pages = {226--231}, year = {1996}}

@article{GaiaCollaboration2018,
  author = {{Gaia Collaboration} and Brown, Anthony G. A. and others},
  title = {Gaia Data Release 2. Summary of the astrometric, photometric, and survey properties},
  journal = {A\&A}, volume = {616}, pages = {A1}, year = {2018}}

@article{GaiaCollaboration2021,
  author = {{Gaia Collaboration} and Brown, Anthony G. A. and others},
  title = {Gaia Early Data Release 3. Summary of the astrometric, photometric, and survey properties},
  journal = {A\&A}, volume = {649}, pages = {A1}, year = {2021}}

@article{GaiaCollaboration2023,
  author = {{Gaia Collaboration} and Vallenari, Antonella and others},
  title = {Gaia Data Release 3. Summary of the astrometric, photometric, and survey properties},
  journal = {A\&A}, volume = {674}, pages = {A1}, year = {2023}}

@misc{GaiaMission2026,
  author = {{ESA Gaia Mission}},
  title = {Gaia mission and data release scenario},
  year = {2026},
  note = {Gaia Data Release 4 expected December 2026; accessed September 2026}}

@article{Gao2020a,
  author = {Gao, Xinhua},
  title = {Detection of tidal tails around the open cluster M 67 using principal component analysis},
  journal = {PASJ}, volume = {72}, pages = {47}, year = {2020}, doi = {10.1093/pasj/psaa031}}

@article{Gao2020b,
  author = {Gao, Xinhua},
  title = {Discovery of Tidal Tails around the Old Open Cluster NGC 2506},
  journal = {ApJ}, volume = {894}, pages = {48}, year = {2020}, doi = {10.3847/1538-4357/ab8560}}

@article{HuntReffert2023,
  author = {Hunt, Emily L. and Reffert, Sabine},
  title = {Improving the open cluster census II. An all-sky cluster catalogue with Gaia DR3},
  journal = {A\&A}, volume = {673}, pages = {A114}, year = {2023}, doi = {10.1051/0004-6361/202346285}}

@article{Jadhav2025,
  author = {Jadhav, Vikrant V. and Risbud, Dhanraj and Kroupa, Pavel and Wu, Wenjie},
  title = {Tidal tails of nearby open clusters II. A review of simulated properties and the reliability of observational catalogues},
  journal = {A\&A}, volume = {704}, pages = {A50}, year = {2025}, doi = {10.1051/0004-6361/202555858}}

@article{Jerabkova2021,
  author = {Jeřábková, Tereza and Boffin, Henri M. J. and Beccari, Giacomo and de Marchi, Guido and de Bruijne, Jos H. J. and Prusti, Timo},
  title = {The 800 pc long tidal tails of the Hyades star cluster. Possible discovery of candidate epicyclic overdensities from an open star cluster},
  journal = {A\&A}, volume = {647}, pages = {A137}, year = {2021}, doi = {10.1051/0004-6361/202039949}}

@article{Just2009,
  author = {Just, Andreas and Berczik, Peter and Petrov, M. I. and Ernst, Andreas},
  title = {Tidal tails of star clusters in the Milky Way},
  journal = {\mnras}, volume = {392}, pages = {969--982}, year = {2009}}

@article{Kos2024,
  author = {Kos, Janez},
  title = {Tidal tails of open clusters},
  journal = {A\&A}, volume = {691}, pages = {A28}, year = {2024}, doi = {10.1051/0004-6361/202449828}}

@article{Kupper2008,
  author = {Küpper, Andreas H. W. and MacLeod, Alexander and Heggie, Douglas C.},
  title = {Tidal tails of star clusters},
  journal = {\mnras}, volume = {387}, pages = {1248--1258}, year = {2008}}

@article{Kupper2010,
  author = {Küpper, Andreas H. W. and Kroupa, Pavel and Baumgardt, Holger and Heggie, Douglas C.},
  title = {Tidal tails of star clusters},
  journal = {\mnras}, volume = {401}, pages = {105--120}, year = {2010}}

@ARTICLE{Li2021,
       author = {{Li}, Yezhang and {Pang}, Xiaoying and {Tang}, Shih-Yun},
        title = "{Evidence of Early-stage Tidal Structures of Open Clusters Revealed by Kinematics with Gaia EDR3}",
      journal = {Research Notes of the American Astronomical Society},
         year = 2021,
        month = jul,
       volume = {5},
       number = {7},
          eid = {173},
        pages = {173},
          doi = {10.3847/2515-5172/ac1688},
archivePrefix = {arXiv},
       eprint = {2107.10853},
 primaryClass = {astro-ph.GA},
       adsurl = {https://ui.adsabs.harvard.edu/abs/2021RNAAS...5..173L}
}

@article{MeingastAlves2019,
  author = {Meingast, Stefan and Alves, João},
  title = {Extended stellar systems in the solar neighborhood. IV. The tidal tails of the Hyades},
  journal = {A\&A}, volume = {621}, pages = {L3}, year = {2019}, doi = {10.1051/0004-6361/201833654}}

@article{Meingast2021,
  author = {Meingast, Stefan and Alves, João and Rottensteiner, Alena},
  title = {Extended stellar systems in the solar neighborhood. V. Discovery of coronae of nearby star clusters},
  journal = {A\&A}, volume = {645}, pages = {A84}, year = {2021}, doi = {10.1051/0004-6361/202038610}}

@ARTICLE{Parul2026,
       author = {{Parul}, Hanna and {Casamiquela}, Laia and {Di Matteo}, Paola and {Ferrone}, Salvatore},
        title = "{Bar-induced deflection of open cluster tidal tails}",
      journal = {arXiv e-prints},
         year = 2026,
        month = jun,
          eid = {arXiv:2606.06248},
        pages = {arXiv:2606.06248},
          doi = {10.48550/arXiv.2606.06248},
archivePrefix = {arXiv},
       eprint = {2606.06248},
 primaryClass = {astro-ph.GA},
       adsurl = {https://ui.adsabs.harvard.edu/abs/2026arXiv260606248P}
}

@article{Risbud2025,
  author = {Risbud, Dhanraj and Jadhav, Vikrant V. and Kroupa, Pavel},
  title = {Tidal tails of nearby open clusters I. Mapping with Gaia DR3},
  journal = {A\&A}, volume = {694}, pages = {A258}, year = {2025}, doi = {10.1051/0004-6361/202453302}}

@article{Roser2019,
  author = {Röser, Siegfried and Schilbach, Elena and Goldman, Bertrand},
  title = {Hyades tidal tails revealed by Gaia DR2},
  journal = {A\&A}, volume = {621}, pages = {L2}, year = {2019}, doi = {10.1051/0004-6361/201834608}}

@article{RoserSchilbach2019,
  author = {Röser, Siegfried and Schilbach, Elena},
  title = {Praesepe (NGC 2632) and its tidal tails},
  journal = {A\&A}, volume = {627}, pages = {A4}, year = {2019}, doi = {10.1051/0004-6361/201935502}}

@article{Sharma2025,
  author = {Sharma, Ira and Jadhav, Vikrant V. and Subramaniam, Annapurni and Wirth, Henriette},
  title = {Tidal tails in open clusters: Morphology, binary fraction, dynamics, and rotation},
  journal = {A\&A}, volume = {704}, pages = {A167}, year = {2025}}

@article{Spitzer1940,
  author = {Spitzer, Lyman},
  title = {Disruption of galactic clusters},
  journal = {\mnras}, volume = {100}, pages = {396--407}, year = {1940}}

@article{Tang2019,
  author = {Tang, Shih-Yun and Pang, Xiaoying and Yuan, Zhen and Chen, W. P. and Hong, Jongsuk and Goldman, Bertrand and Just, Andreas and Shukirgaliyev, Bekdaulet and Lin, Chien-Cheng},
  title = {Discovery of Tidal Tails in Disrupting Open Clusters: Coma Berenices and a Neighbor Stellar Group},
  journal = {ApJ}, volume = {877}, pages = {12}, year = {2019}}

@article{Tarricq2022,
  author = {Tarricq, Yoann and Soubiran, Corinne and Casamiquela, Laia and Castro-Ginard, Adriano and Olivares, Javier and Miret-Roig, Núria and Galli, Pablo A. B.},
  title = {Structural parameters of 389 local open clusters},
  journal = {A\&A}, volume = {659}, pages = {A59}, year = {2022}, doi = {10.1051/0004-6361/202142186}}

@article{Weis2025,
  author = {Weis, L. and West, C. and Just, A. and Berczik, P. and Ishchenko, M. and Röser, S. and Schilbach, E. and Shukirgaliyev, B.},
  title = {Dynamical model of Praesepe and its tidal tails},
  journal = {A\&A}, volume = {703}, pages = {A47}, year = {2025}, doi = {10.1051/0004-6361/202451650}}

@ARTICLE{Ye2021,
       author = {{Ye}, Xianhao and {Zhao}, Jinkun and {Zhang}, Jiajun and {Yang}, Yong and {Zhao}, Gang},
        title = "{Extended Tidal Tails of IC 4756 Detected by Gaia EDR3}",
      journal = {\aj},
         year = 2021,
        month = oct,
       volume = {162},
       number = {4},
          eid = {171},
        pages = {171},
          doi = {10.3847/1538-3881/ac1f1f},
archivePrefix = {arXiv},
       eprint = {2110.08104},
 primaryClass = {astro-ph.GA},
       adsurl = {https://ui.adsabs.harvard.edu/abs/2021AJ....162..171Y}
}

@article{Zhang2020,
  author = {Zhang, Jiajun and Zhao, Jinkun and Oswalt, Terry D. and Fang, Xiangsong and Zhao, Gang},
  title = {Gaia DR2 study of the nearby open cluster Blanco 1},
  journal = {ApJ}, volume = {889}, pages = {99}, year = {2020}}

@article{Zhao2026,
  author = {Zhao, Wenchang and Chen, Jing and Tang, Yujun and He, Xuran and Deng, Zhengke and Huang, Lishuai and Fu, Lin and Chi, Huanbin and Zhang, Su},
  title = {Tidal tails and mass segregation in nearby open clusters with Gaia DR3},
  journal = {A\&A}, volume = {711}, pages = {A188}, year = {2026}, doi = {10.1051/0004-6361/202660529}}

\appendix

\section{Literature comparison tables}

\begin{table*}
\caption{Comparison of principal methodological families used in open-cluster membership and tidal-tail studies.}
\label{tab:methods}
\centering
\small
\begin{tabular}{p{2.4cm}p{3.6cm}p{4.0cm}p{4.0cm}}
\hline
Method & Core idea & Main strength for tidal-tail searches & Main limitation\\
\hline
Convergent point & Select stars whose proper motions are consistent with the cluster bulk motion & Simple, physically motivated, effective for nearby moving systems & Sensitive to distance and velocity errors; co-moving field contamination\\
DBSCAN & Find density-connected groups above a fixed density scale & Recovers arbitrary-shaped structures & Sensitive to density threshold and field gradients\\
HDBSCAN & Hierarchical density estimation over multiple density levels & Handles varying density and non-spherical structures & Still depends on metric/scaling and cluster-selection parameters\\
GMM & Model cluster and field as probability distributions & Provides membership probabilities and uncertainty framework & Gaussian components can suppress curved or diffuse tails\\
kNN / ML-MOC & Classify stars from local similarity to a reference population & Flexible and scalable membership refinement & Depends on reference/training population and local density\\
PCA + GMM & Reduce dimensionality before probabilistic clustering & Useful for high-dimensional Gaia data & Projection can remove physically relevant low-density structure\\
Orbit integration & Test whether candidate stars are dynamically compatible with cluster orbit & Direct physical interpretation & Depends on Galactic potential and initial conditions\\
N-body matching & Compare data with simulated dissolution models & Predicts morphology, escape, and substructure jointly & Computationally expensive and model dependent\\
\hline
\end{tabular}
\end{table*}

\begin{table*}
\caption{Representative open clusters and the principal evidence discussed in the literature. ``Extended'' denotes a cluster-associated population beyond a conventional structural radius; ``tail'' is reserved for a structure with a stronger dynamical interpretation.}
\label{tab:clusters}
\centering
\small
\begin{tabular}{p{2.4cm}p{3.6cm}p{4.0cm}p{4.0cm}}
\hline
Cluster & Main references & Reported evidence & Interpretation in this review\\
\hline
Hyades & \citet{Roser2019}; \citet{MeingastAlves2019}; \citet{Jerabkova2021} & Leading/trailing tails; several 100 pc & Benchmark tail detection\\
Praesepe & \citet{RoserSchilbach2019}; \citet{Meingast2021}; \citet{Kos2024} & Extended population; detailed structural model & Strong extended/tidal structure\\
Coma Berenices & \citet{Tang2019}; \citet{Tarricq2022}; \citet{Kos2024} & $\sim$50 pc tails in early work; later systematic detections & Strong extended population; morphology method-dependent\\
Blanco 1 & \citet{Zhang2020}; \citet{Meingast2021}; \citet{Li2021} & $\sim$50--60 pc tails & Strong candidate tail; young-cluster caveat\\
NGC 2516 & \citet{Li2021}; \citet{Tarricq2022} & Early-stage extensions & Candidate early tidal disruption\\
Pleiades & \citet{Li2021}; \citet{Tarricq2022} & Kinematic evidence for early disruption & Extended population; not a mature tail benchmark\\
M\,67 & \citet{Gao2020a}; \citet{Kos2024}; \citet{Sharma2025} & Extra-tidal population and tails & Strong extended structure\\
NGC 752 & \citet{Boffin2022}; \citet{Tarricq2022} & Long asymmetric tails & Strong tail detection\\
IC 4756 & \citet{Ye2021} & Two tails to $\sim$180 pc & Strong candidate tail\\
NGC 7092 & \citet{Meingast2021}; \citet{Tarricq2022} & Extended/tidal structure & Candidate tail\\
COIN-Gaia 13 & \citet{Bai2022} & $\sim$270 pc elongated structure & Extended population; tidal interpretation uncertain\\
BH 164 & \citet{Sharma2025} & 40--100 pc tail-scale structure & Recent candidate tail\\
Alessi 2 & \citet{Sharma2025} & 40--100 pc tail-scale structure & Recent candidate tail\\
NGC 2281 & \citet{Sharma2025} & 40--100 pc tail-scale structure & Recent candidate tail\\
NGC 2354 & \citet{Sharma2025} & 40--100 pc tail-scale structure & Recent candidate tail\\
NGC 2972 & \citet{Zhao2026} & Gaia DR3 tail structure & Recent homogeneous survey\\
\hline
\end{tabular}
\end{table*}

\end{document}